\documentclass[%
 aip,
 amsmath,amssymb,
 reprint,%
]{revtex4-1}

\usepackage{graphicx}
\usepackage{dcolumn}
\usepackage{bm}

\usepackage[utf8]{inputenc}
\usepackage[T1]{fontenc}
\usepackage{mathptmx}
\usepackage{etoolbox}
\usepackage[per-mode=repeated-symbol]{siunitx}
\usepackage{array}
\usepackage[mathscr]{euscript}
\usepackage[dvipsnames]{xcolor}

\makeatletter
\def\@email#1#2{%
 \endgroup
 \patchcmd{\titleblock@produce}
  {\frontmatter@RRAPformat}
  {\frontmatter@RRAPformat{\produce@RRAP{*#1\href{mailto:#2}{#2}}}\frontmatter@RRAPformat}
  {}{}
}%
\makeatother
\begin{document}

\preprint{AIP/123-QED}

\title[Influence of the laser pulse duration in high-order harmonic generation]{Influence of the laser pulse duration in high-order harmonic generation}
\author{S. Westerberg\textsuperscript{*}}
 \email[]{saga.westerberg@fysik.lth.se, anne-lise.viotti@fysik.lth.se}
 \affiliation{Department of Physics, Lund University, P.O. Box 118, SE-22100 Lund, Sweden}

\author{M. Redon}
 \affiliation{Department of Physics, Lund University, P.O. Box 118, SE-22100 Lund, Sweden}
\author{A.-K. Raab}
 \affiliation{Department of Physics, Lund University, P.O. Box 118, SE-22100 Lund, Sweden}
\author{G. Beaufort}
 \affiliation{Department of Physics, Lund University, P.O. Box 118, SE-22100 Lund, Sweden}
\author{M. Arias Velasco}
 \affiliation{Institute for Photonics and Nanotechnologies (IFN), Consiglio Nazionale delle Ricerche (CNR), Piazza Leonardo da Vinci 32, 20133 Milano, Italy}
 \affiliation{Department of Physics, Politecnico di Milano, Piazza Leonardo da Vinci 32, 20133 Milano, Italy}
\author{C. Guo}
 \affiliation{Department of Physics, Lund University, P.O. Box 118, SE-22100 Lund, Sweden}
 \author{I. Sytcevich}
 \affiliation{Department of Physics, Lund University, P.O. Box 118, SE-22100 Lund, Sweden}
 \author{R. Weissenbilder}
 \affiliation{Department of Physics, Lund University, P.O. Box 118, SE-22100 Lund, Sweden}
 \author{D. O'Dwyer}
 \affiliation{ASML Research, ASML Netherlands B.V., 5504 DR Veldhoven, The Netherlands}
  \author{P. Smorenburg}
 \affiliation{ASML Research, ASML Netherlands B.V., 5504 DR Veldhoven, The Netherlands}
\author{C. L. Arnold}
 \affiliation{Department of Physics, Lund University, P.O. Box 118, SE-22100 Lund, Sweden}
 \author{A. L'Huillier}
 \affiliation{Department of Physics, Lund University, P.O. Box 118, SE-22100 Lund, Sweden}
\author{A.-L. Viotti\textsuperscript{*}}
 \affiliation{Department of Physics, Lund University, P.O. Box 118, SE-22100 Lund, Sweden}
\date{\today}

\begin{abstract}
High-order harmonic generation (HHG) in gases has been studied for almost \mbox{40 years} in many different conditions, varying the laser wavelength, intensity, focusing geometry, target design, gas species, etc. However, no systematic investigation of the effect of the pulse duration has been performed in spite of its expected impact on phase-matching of the high-order harmonics.
Here, we develop a compact post-compression method based on a bulk multi-pass cell enabling tunable Fourier-limited pulse durations. We examine the HHG yield as a function of the pulse duration, ranging from $\SI{42}{\femto\second}$ to $\SI{180}{\femto\second}$, while maintaining identical focusing conditions and generating medium. Our findings reveal that, for a given intensity, there exists an optimum pulse duration - not necessarily the shortest - that maximizes conversion efficiency. This optimum pulse duration increases as the intensity decreases.
The experimental results are corroborated by numerical simulations, which show the dependence of HHG yield on the duration and peak intensity of the driving laser and underscore the importance of the interplay between light-matter interaction and phase-matching in the non-linear medium. Our conclusion explains why HHG could be demonstrated in 1988 with pulses as long as $\SI{40}{\pico\second}$ and intensities of just a few \mbox{$10^{13}$ W/cm${^2}$}.

\end{abstract}

\maketitle

\section{Introduction}

High-order harmonic generation (HHG) in gases \cite{McPherson87} , which takes place when atoms are exposed to intense laser pulses, provides coherent extreme ultraviolet (XUV) radiation with pulse durations down to the attosecond regime in a table-top environment \cite{Popmintchev:2010}. The short duration enables high temporal resolution in pump-probe experiments \cite{Kretschmar:22}, while the short wavelength allows high spatial resolution in imaging \cite{porterSoftXrayNovel2023}. Nevertheless, many applications, such as coincidence spectroscopy \cite{Mikaelsson+2021+117+128} or coherent imaging \cite{zürch2014real}, require high repetition rates and large XUV flux. The latter is challenging due to the inherently low conversion efficiency (CE) of the HHG process ($10^{-7} - 10^{-5}$) \cite{weissenbilderHow2022,Tempea:2000}. 

The repetition rate of the XUV source follows that of the driving laser. For many years, Titanium:Sapphire systems with pulse durations down to the few-cycle regime \cite{PhysRevLett.76.752, PM.Paul-obsAsTrain} have been the workhorse of HHG. However, their repetition rates are limited to just a few kHz. Ytterbium (Yb)-based sources can operate at much higher repetition rates, up to the MHz regime \cite{limpert2002high}, are power scalable and come at a reasonable cost. Yet, Yb amplifiers deliver longer pulse durations in the range of hundreds of fs to $\SI{1}{\pico\second}$ \cite{Fattahi:14}. To overcome this, post-compression via self-phase modulation (SPM) is commonly used with Yb sources \cite{nagy_high-energy_2021}. It can be implemented with different methods, for example, hollow-core fibers \cite{Fan:21}, multiple-plate continuum \cite{Lu:14} or multi-pass cells (MPCs) \cite{schulte_nonlinear_2016}. The MPC scheme combines efficiency \cite{pfaffNonlinear2023} with compactness \cite{haritonSpectral2023,omarSpectral2023} and allows relatively simple tuning of the spectral broadening \cite{viottiFLASH2023}. 

In parallel, a lot of effort has been made to increase the CE of HHG using, for example, multiple colors \cite{raab2024xuv}, short laser pulses \cite{schmidt2012high}, optimization of the medium properties or the focusing geometry \cite{heyl2017introduction}.
Some of these parameters affect the single-atom response, while others influence how the radiation is built up in the macroscopic medium (i.e. phase-matching). 
In most experimental conditions, phase-matching is achieved when a certain ionization degree is reached in the medium, at which the phase velocities of the fundamental and harmonic fields are the same \cite{weissenbilderHow2022}. This ionization degree, typically of the order of a few percent, is reached at different laser intensities for different pulse durations. This implies that the pulse duration of the fundamental field will affect the harmonic flux and therefore the CE. 
    
Tempea and Brabec \cite{Tempea:2000} study numerically the generation of high-order harmonics in several gases as a function of pulse duration. In general, they find increasing CEs with decreasing pulse duration, for the same laser intensity. Hädrich et al. \cite{Hädrich_2016}, using the one-dimensional theoretical model of Constant et al. \cite{Constant:1999}, report a rapid increase of the 25$^{\text{th}}$ harmonic signal, generated in xenon, as the driver's duration is decreased. This study is performed at a fixed ionization degree in the medium. Experimentally, multiple groups have used few-cycle pulses to achieve high XUV yields and high photon energies \cite{Johnson:2018,schmidt2012high}. However, no systematic pulse duration investigation employing the same laser source has been performed so far due to experimental complexity. 

One simple way to vary the pulse duration is to chirp the fundamental field. But, this also changes the light-matter interaction beyond just varying the pulse duration, notably by affecting the single-atom response \cite{ChirpHHG_sam}. This motivates a study where the duration of the driving field is tuned while keeping it Fourier transform limited. Moreover, to understand the influence of the duration, other parameters need to be kept constant, such as the focusing geometry, medium length, and pressure. As the ionization degree in the medium depends both on the laser intensity and the pulse duration, a systematic study of the influence of the pulse duration on HHG therefore requires to vary both parameters.

With this in mind, we have developed and characterized a compact pulse post-compression setup for an Yb system, based on a bulk MPC. The duration of the compressed output pulses can be tuned in the range \qtyrange{42}{80}{\femto \second} without altering any spatial properties. We experimentally study the generation of high-order harmonics in argon as a function of pulse duration, using the tunable pulses from the MPC as well as the 
non-compressed pulse ($\SI{180}{\femto\second}$).
We also simulate HHG by numerically solving the time-dependent Schr\"odinger equation and propagation equations \cite{PhysRevA.46.2778}.
Our results lead to a better understanding of the HHG process and highlight the influence of the pulse duration for phase-matching. In particular, we find that, for a given laser intensity, there is an optimum pulse duration, not necessarily the shortest, that maximizes the CE. This optimum pulse duration ($\tau_\textrm{opt}$) is such that a certain ionization degree can be reached in the medium, thus achieving phase-matching. The required pulse duration increases as the intensity decreases. Extrapolating to very long pulses, we can now understand why HHG could be observed with pulses as long as $\SI{40}{\pico\second}$ and intensities much below $10^{14}$ W/cm${^2}$, as in the first demonstration~\cite{MFerray_1988} leading to a Nobel prize in 2023. Using short laser pulses and high laser intensities leads, however, to higher CEs. The maximum CE that can be obtained is found to vary approximately as $\tau_\textrm{opt}^{-0.5}$.

\section{Methods}

\subsection{Pulse post-compression in a bulk multi-pass cell}

The experimental frontend consists of an Yb laser (Light Conversion, Pharos) with a $\SI{1030}{\nano\meter}$ central wavelength, a full width at half maximum (FWHM) pulse duration of $\SI{180}{\femto\second}$, a repetition rate of $\SI{20}{\kilo\hertz}$ and pulse energies up to $\SI{700}{\micro\joule}$. 
To generate pulses with different durations, the post-compression scheme needs to be tunable. At the same time, to isolate the effect of pulse duration on the HHG yield, the focusing geometry for HHG and the parameters of the gas target must remain unchanged. This puts additional requirements on the post-compression setup, as the change in pulse duration should be performed with minimum realignment and without significantly modifying the spatial profile. While there are many available pulse post-compression techniques \cite{nagy_high-energy_2021}, the requirement of high efficiency (typically $\SI{85}{\percent}-\SI{95}{\percent}$ \cite{grebing2020kilowatt}), compactness and tunable compression motivates the choice of MPC. 

A conventional MPC setup consists of two focusing mirrors forming an optical resonator, in which a non-linear medium is placed \cite{hanna_nonlinear_2021, viottiMultipass2022}. Spectral broadening is progressively achieved through SPM in the Kerr medium over multiple passes. During this process, a spectral phase, mainly quadratic, is accumulated and needs to be compensated for in order to achieve compression. This is usually performed with dispersive mirrors or grating compressors. In an MPC, the beam enters the resonator off-axis, and propagates between the two mirrors according to the Herriott re-entrant condition \cite{herriott1964off}, forming a circular pattern on the mirrors. After a certain number of round-trips, the beam is extracted from the cavity. As this forms an optical resonator, the beam inside the MPC must be mode-matched to its eigenmode, ensuring that the beam sizes on the mirrors and in the non-linear material remain unchanged during propagation. The non-linear medium can be either gas or bulk material. Gas-filled MPCs are typically employed when the peak power of the laser exceeds $\SI{1}{\giga\watt}$ \cite{pfaffNonlinear2023} but require the use of vacuum chambers. At peak powers in the range of $\SI{1}{\mega\watt}$ to $\SI{1}{\giga\watt}$, bulk-based  MPCs, for example with glass plates, are more common, leading to more compact setups \cite{omar_hybrid_2024,viottiFewcycle2023}. Bulk materials have a high non-linear refractive index, compared to noble gases, and a large spectral broadening can be obtained at relatively low intensity. The most common material used in bulk MPCs is fused silica, with a non-linear index of \mbox{$3\times 10^{-16}$ cm${^2}$/W} at $\SI{1030}{\nano\meter}$ \cite{milam1998review}.

Using high peak powers in a bulk MPC can bring several challenges. First, the ionization threshold of the bulk material may be exceeded, leading to permanent damage. Second, beam degradation may occur when operating far above the critical power for self-focusing. Experiments have shown that this can be mitigated in bulk MPCs by employing thin non-linear material placed out of focus \cite{raabMultigigawatt2022}, so that the beam exits the medium before it collapses or suffers significant degradation. Third, lensing in the non-linear material due to self-focusing can unmatch the beam from the cavity eigenmode, causing the beam sizes on the mirrors and position of the focus, to change from pass to pass. This can lead to damage both in the non-linear medium and on the cavity mirrors, and the amount of non-linearity per pass may vary. Thus, mode-matching should account for lensing effects arising from large peak powers \cite{hanna_nonlinear_2021-1,seidelFactor2022}. In our experiment, the peak power is up to 300 times the critical power of fused silica, which is $\SI{4.3}{\mega\watt}$ at $\SI{1030}{\nano\meter}$ \cite{fibich_critical_2000}. Long propagation within the medium would lead to catastrophic collapse or small-scale self-focusing, ruining beam quality and transmission \cite{Martyanov_Ginzburg_Balakin_Skobelev_Silin_Kochetkov_Yakovlev_Kuzmin_Mironov_Shaikin_et_al._2023}. By keeping the medium short, beam collapse is avoided. Finally, parasitic non-linear effects in air can cause an asymmetric broadening with a complicated spectral phase that cannot be easily corrected. Those effects are avoided by removing the air in the center of the cavity, where the intensity is the highest.

\subsection{Experimental setup}

\begin{figure}[htbp]
\centering
\includegraphics[width=1\linewidth]{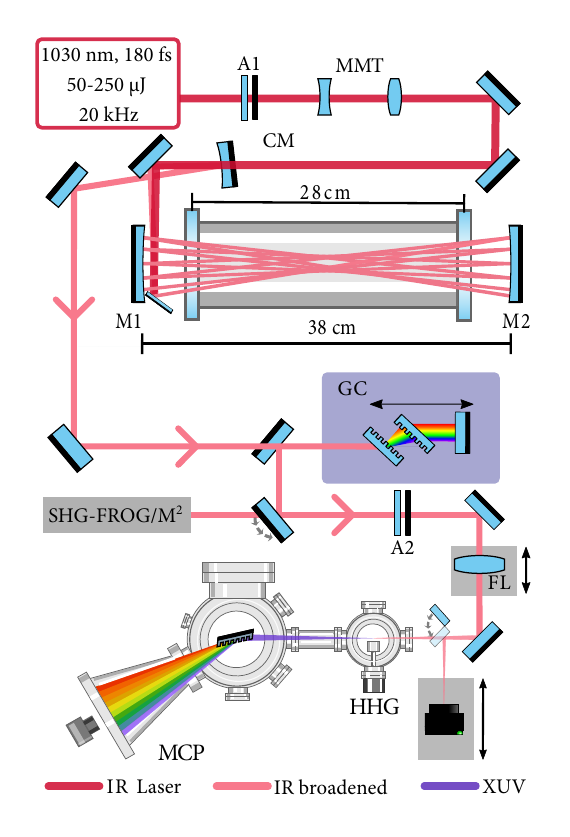}
\caption{The experimental setup. A1 and A2: variable attenuators. MMT: mode-matching telescope. M1 and M2: MPC cavity mirrors. CM: collimating mirror. GC: grating compressor. FL: focusing lens. HHG: vacuum chamber with the gas target. MCP: micro-channel plate. SHG-FROG: second harmonic generation frequency-resolved optical gating.}
\label{fig:Setup}
\end{figure}

The experimental setup is shown in Fig.~\ref{fig:Setup}. At the exit of the laser, an attenuator, A1, consisting of a half-wave plate and a thin film polarizer, controls the input pulse energy, ranging from $\SI{50}{\micro\joule}$ to $\SI{250}{\micro\joule}$. 
The beam is then sent to a lens telescope, MMT, which mode-matches the beam to the first eigenmode of a Herriott-type MPC. The MPC mirrors, M1 and M2, have a radius of curvature (ROC) of $\SI{200}{\milli\meter}$, the cell length is $\SI{382}{\milli\meter}$ and the beam propagates a total of seven round-trips. Placed symmetrically around the center of the MPC and separated by $\SI{280}{\milli\meter}$, two $\SI{1}{\milli\meter}$-thick anti-reflection coated fused silica plates act as a Kerr medium. The two fused silica plates constitute the windows of a small cell, evacuated to around $\SI{5}{\milli\bar}$. The beam enters and exits the MPC via the same coupling mirror, and is collimated using a curved mirror, CM, with $\SI{1500}{\milli\meter}$ ROC. The spectrally broadened pulse is sent to a  double pass grating compressor, GC, consisting of two transmission gratings with $1000$~lines/mm. The second grating and the retro-reflector are mounted on a motorized translation stage, allowing precise adjustment of the grating distance and therefore of the dispersion.

\subsection{Characterization of the MPC output and strategy for pulse duration tuning}

\begin{figure}
\centering
\includegraphics[width=1\linewidth]{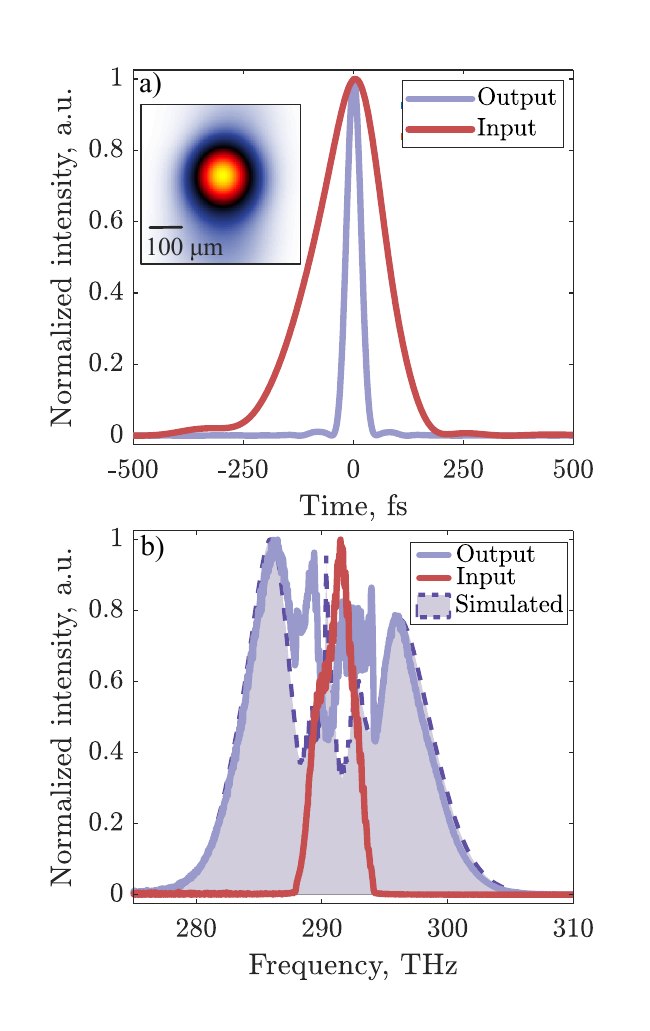}
\caption{a) Retrieved temporal profile of the input (red) and compressed output pulse (purple) at $\SI{250}{\micro\joule}$, measured with SHG-FROG. The FWHM durations are $\SI{180}{\femto\second}$ and $\SI{42}{\femto\second}$ for the input and compressed pulses, respectively. The inset shows the output beam profile in the focus. b) Measured input (red) and output (purple) spectra together with the simulated spectrum (shaded) obtained with SISYFOS.}
\label{fig:Characterization}
\end{figure}

Second harmonic generation frequency-resolved optical gating (SHG-FROG) is used throughout the study to characterize the different driver pulse durations. Figure~\ref{fig:Characterization} a) presents the temporal profiles of the input and compressed pulses at the maximum input pulse energy of $\SI{250}{\micro\joule}$. The duration of the compressed pulse is $\SI{42}{\femto\second}$ FWHM, with a Fourier Transform Limit (FTL) of $\SI{39}{\femto\second}$. The corresponding measured input and output spectra are shown in Fig.~\ref{fig:Characterization} b).

Several approaches can be taken to tune the pulse duration: changing the number of round-trips, the distance between the fused silica plates, or the input energy. Adjusting the number of round-trips requires a change in the length of the MPC, which alters mode-matching and alignment. The collimation of the beam is also modified, which subsequently affects phase-matching of the high-order harmonics, and thus complicates the comparison between results obtained at different pulse durations. The second approach, moving the fused silica plates, also requires the non-linear mode-matching to be adjusted. Additionally, when the plates are placed closer to the focus the quality of the beam profile and the transmission of the MPC suffers, due to self-focusing effects. The third approach, varying the input energy, allows minimal realignment as the beam path and beam profile are not energy-dependent. Moreover, a given non-linear mode-matching can be realized for a relatively wide range of pulse energies. With this tuning scheme, the longer pulses contain less energy than the shorter ones, and therefore the maximum achievable peak intensities are lower. While this is a limit in the experiment, pulse duration studies can be performed sequentially within the same day, under similar conditions. 

\begin{figure}
\centering
    \includegraphics[width=1\linewidth]{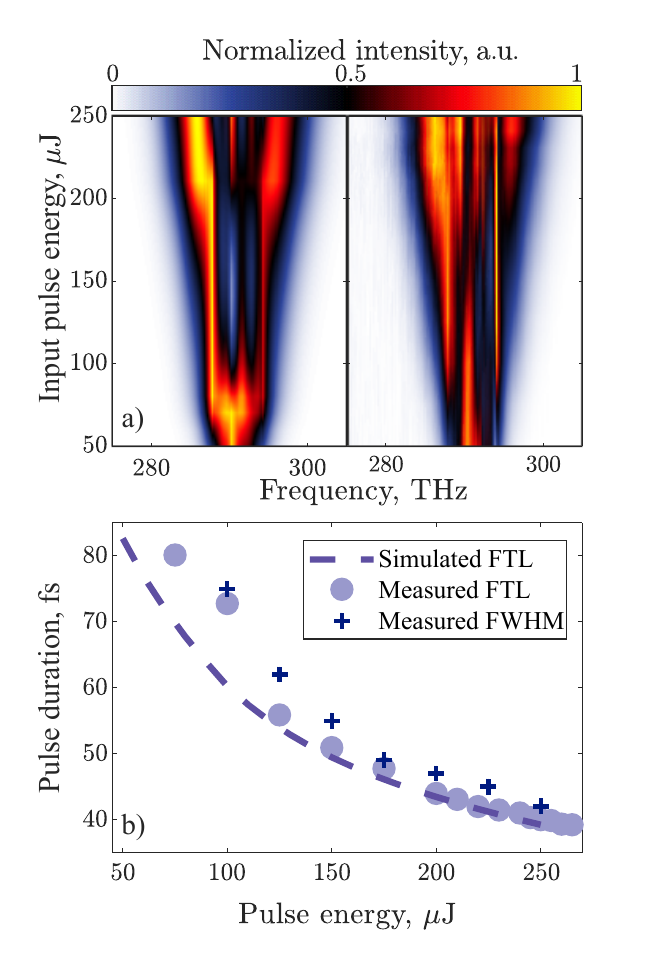}
    \caption{a) Simulated (left) and measured (right) spectra as a function of the MPC input pulse energy. b) Simulated (dashed line) and measured (dots) FTL FWHM of the MPC output pulse duration for different input pulse energies. The crosses show the retrieved durations from SHG-FROG measurements.}
    \label{fig:Tuning}
\end{figure}

We simulate the propagation of the pulse in the MPC using the  SISYFOS code \cite{Sisyfos}. The input pulse is obtained from the measured laser spectrum, and the MPC parameters are those of the experiment. The reflectivity of the mirrors is defined from the coating specification and we account for the measured $\text{M}^2$ to determine the peak intensity. Figure~\ref{fig:Characterization} b) displays the simulated spectrum using a $\SI{250}{\micro\joule}$ input pulse energy, which is in good agreement with the measured spectrum. By iterating the simulation with input pulse energies from $\SI{50}{\micro\joule}$ to $\SI{250}{\micro\joule}$, we obtain the spectra shown on the left in Fig.~\ref{fig:Tuning} a), in agreement with the measured spectra (right). The FTL pulse durations for simulated and experimental spectra are compared  with the retrieved pulse durations measured with SHG-FROG in Fig.~\ref{fig:Tuning} b). The deviation observed between simulated and measured FTLs for longer pulses can be explained by the difference in spectral distribution in Fig.~\ref{fig:Tuning} a). The simulation assumes a flat spectral phase and a perfect Gaussian beam at the input.

The transmission of the MPC is $>\SI{94}{\percent}$ and the grating compressor ($-6144$~fs${^2}$/mm) has a transmission $>\SI{90}{\percent}$. This results in an overall power efficiency of the compression system $>\SI{84}{\percent}$. The peak power obtained by compressing the $\SI{250}{\micro\joule}$ input pulses to $\SI{42}{\femto\second}$ is $\SI{4.8}{\giga\watt}$. The input beam has an $\text{M}^2$ of $1.1\times 1.1$, and the $\text{M}^2$ of the output beam is measured to $1.1\times 1.3$, implying good focusability. The focused output beam profile is shown as an inset in Fig.~\ref{fig:Characterization} a). 

\subsection{High-order harmonic generation: experiments}

After compression, the pulses are sent to a motorized attenuator, A2, which controls the pulse energy. A $\SI{200}{\milli\meter}$ focusing lens, FL, placed on a motorized translation stage, focuses the beam into a $\SI{100}{\micro\meter}$-long argon gas jet from a nozzle with a backing pressure of $\SI{2.5}{\bar}$. The Rayleigh length is estimated to $\SI{2.7}{\milli\meter}$. The XUV radiation originating from HHG is sent to a spectrometer consisting of a curved XUV grating (ROC $\SI{5649}{\milli\meter}$, $600$ lines/mm) and a micro-channel plate (MCP) coupled to a phosphor screen imaged by a camera.

The generating field is characterized temporally via SHG-FROG as close to the chamber as possible. A small portion of the beam is extracted before entering the vacuum chamber to monitor the focus size, shape and position with respect to the gas target (see Fig.~\ref{fig:Setup}). The power is measured just before the chamber and the motorized attenuator is calibrated accordingly. We neglect possible losses from the anti-reflection coated chamber window. The peak power is obtained after normalizing the retrieved temporal profiles by the pulse energy, and the peak intensity is estimated after normalizing the spatial profile in the focus by the peak power. 

For each pulse duration, the total harmonic signal is optimized. 
The dispersion is minimized by tuning the grating compressor, thus ensuring compressed pulses in the generation. The backing pressure is kept constant. The XUV spectra are then recorded as a function of the fundamental pulse energy in the chamber. To vary the pulse duration, the half-wave plate of the first attenuator (A1) is turned, changing the input energy in the MPC from $\SI{100}{\micro\joule}$ to $\SI{225}{\micro\joule}$, with a $\SI{25}{\micro\joule}$ spacing, resulting in compressed pulses of $\SI{75}{\femto\second}$, $\SI{65}{\femto\second}$, $\SI{57}{\femto\second}$, $\SI{49}{\femto\second}$ and $\SI{45}{\femto\second}$.

We study the effect of the pulse duration, both on the total HHG yield and on individual harmonics, from 19 to 37 ($\SI{20}{\electronvolt}$ to $\SI{45}{\electronvolt}$), which is the detection range of our spectrometer. HHG strongly depends on pulse duration and intensity, leading to varying number of counts, which require different MCP voltages to keep a good signal-to-noise ratio. A measurement series taken with the same input field and different MCP voltage values is used to calibrate the HHG spectra.

\subsection{High-order harmonic generation: simulations}

We perform simulations \cite{weissenbilderHow2022,PhysRevA.46.2778} to understand the influence of the pulse duration on harmonic yield and conversion efficiency. The single-atom response is obtained by solving the time-dependent Schr\"odinger equation (TDSE) for an argon atom within the single-active electron approximation. The propagation of the fundamental and the harmonic fields in the non-linear medium is calculated using the paraxial and slowly varying envelope approximations. We assume Fourier-limited pulses, with a Gaussian temporal profile, and a Gaussian beam with a waist of $\SI{30}{\micro\meter}$ corresponding to a Rayleigh length of $\SI{2.7}{\milli\meter}$. The length of the medium is $\SI{100}{\micro\meter}$ and a pressure of $\SI{1}{\bar}$ is assumed at the generation point. To match the experimental conditions, we only consider the harmonic emission within a half-angle of $\SI{10}{\milli rad}$.

\section{Results and Discussion}

\subsection{Basic principles of HHG and effect of pulse duration on HHG yield}

To facilitate the discussion of the experimental and simulated results presented below, we first summarize the fundamental physics governing the generation of high-order harmonics, outlining the influence of the laser pulse duration. 
HHG requires both that the light-matter interaction is in the strong-field regime and that the XUV emission from a large number of atoms in the macroscopic medium adds in phase (phase-matching). The first condition implies a minimum intensity so that tunnel-ionized electrons reach sufficient kinetic energy on their trajectories in the laser field to generate harmonics. The maximum intensity to which an atom can be exposed before the single-atom response saturates, traditionally called the ``saturation'' intensity, is such that the ionization probability becomes significant, typically more than $\SI{50}{\percent}$. This effect depends on the pulse duration and occurs at increasing intensity with decreasing pulse duration. The high intensity that an atom can experience when using a short laser pulse leads to a high HHG yield and to a plateau extending to high photon energies. However, for a given intensity such that the ionization probability remains small, the single-atom response depends only weakly on the pulse duration. The second condition leads to a maximum intensity, above which phase-matching is not possible. 

To achieve phase-matching, the phase-mismatch between the fundamental and $q$th harmonic fields, \sloppy ${\Delta k = qk_1-k_q}$, should be approximately equal to 0. Here, $k_1$ and $k_q$ are the wave-vectors for the fundamental and $q$th harmonic fields, respectively. Four terms contribute to $\Delta k$: the neutral atom dispersion $(\Delta k_{\text{at}}>0)$, the dispersion due to free electrons $(\Delta k_{\text{fe}}<0)$, the effect of focusing $(\Delta k_{\text{foc}}<0)$ and the effect of the electron trajectories in the continuum ($\Delta k_{\text{i}}\mathop{\lessgtr}0$) \cite{weissenbilderHow2022}. In our experimental conditions such that the medium is short relative to the Rayleigh length and is centered at the focus of the fundamental, the last contribution is small. 
The free-electron contribution $\Delta k_{\text{fe}}$ is proportional to the ionization degree in the medium, which depends on the laser intensity and pulse duration. The ionization degree at which phase-matching is achieved, is such that $\Delta k_{\text{at}}+\Delta k_{\text{fe}}+\Delta k_{\text{foc}} \approx 0$. To obtain optimum phase-matching, this \textit{critical} ionization degree should be reached near the peak of the pulse, which defines a \textit{critical intensity} for a given pulse duration.

In argon driven by $\SI{1030}{\nano\meter}$ radiation, the critical ionization degree is in the range of $\SI{2}{\percent}$ to $\SI{4}{\percent}$, depending on the harmonic order $q$.
The critical intensity is, therefore, smaller than the saturation intensity mentioned previously, which implies that for efficient HHG, the single-atom response can be considered independent of the pulse duration. Phase-matching, however, strongly depends on the pulse duration, following the variation of the critical ionization degree (and critical intensity). The harmonic yield $N_q$ is an integral of the square of the harmonic field over time and space and can be expressed as the product of a contribution from the single-atom response ($A_q$), a contribution from phase-matching ($F_q$) and the laser pulse duration $\tau$. The quantities $A_q$ and $F_q$ are intensity-dependent, while only $F_q$ depends on the fundamental pulse duration, 
\begin{equation}
    N_q(I,\tau) \approx A_q(I)F_q(I,\tau)\tau.
\end{equation}
The CE ($C_q$), defined as the ratio of the pulse energies of the harmonic and fundamental fields, is proportional to
\begin{equation}
    C_q(I,\tau) \propto \frac{A_q(I)F_q(I,\tau)}{I}.
\end{equation}
Based on these simple arguments, we expect that for a given, fixed intensity, the CE should reflect the dependence of $F_q(I,\tau)$ with a maximum at a certain pulse duration. A short pulse duration leads to phase-matching being realized at a high laser intensity, thus to a strong single-atom response and high harmonic yield. A longer pulse duration leads to phase-matching being realized at lower laser intensity, thus to a smaller contribution of the single-atom response, only partly compensated by the longer pulse length, i.e. the additional number of cycles available for the generation.    

When $|\Delta k_{\text{fe}}|$ is small compared to $|\Delta k_{\text{at}}|$ and $|\Delta k_{\text{foc}}|$, phase-matching does not depend as much as previously on the ionization degree and therefore on the pulse duration. In this case, the harmonic yield increases linearly with the pulse duration, and the CE does not depend on the pulse duration. High intensities can be reached using short pulses, leading to optimized CEs for the shortest possible duration. We will only comment briefly this regime, which is less explored and was not that of our experiments.

\begin{figure}
    \centering
    \includegraphics[width=1\linewidth]{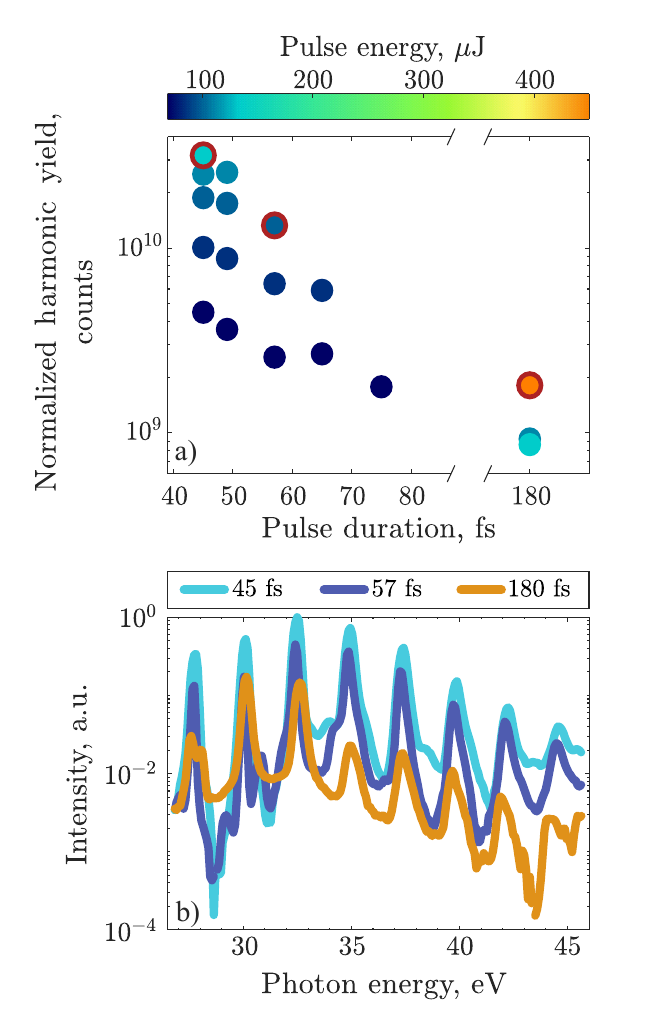}
    \caption{a) Harmonic yield normalized by the pulse energy of the generating field, integrated over the harmonic spectrum, as a function of pulse duration for different pulse energies indicated in color. b) Harmonic spectra corresponding to the three pulse energies and pulse durations highlighted by the red circles in a). All spectra are normalized by the maximum signal of the $\SI{45}{\femto\second}$ case.}
    \label{fig:Result_1}
\end{figure}

\begin{figure*}
    \includegraphics[width=1\textwidth]{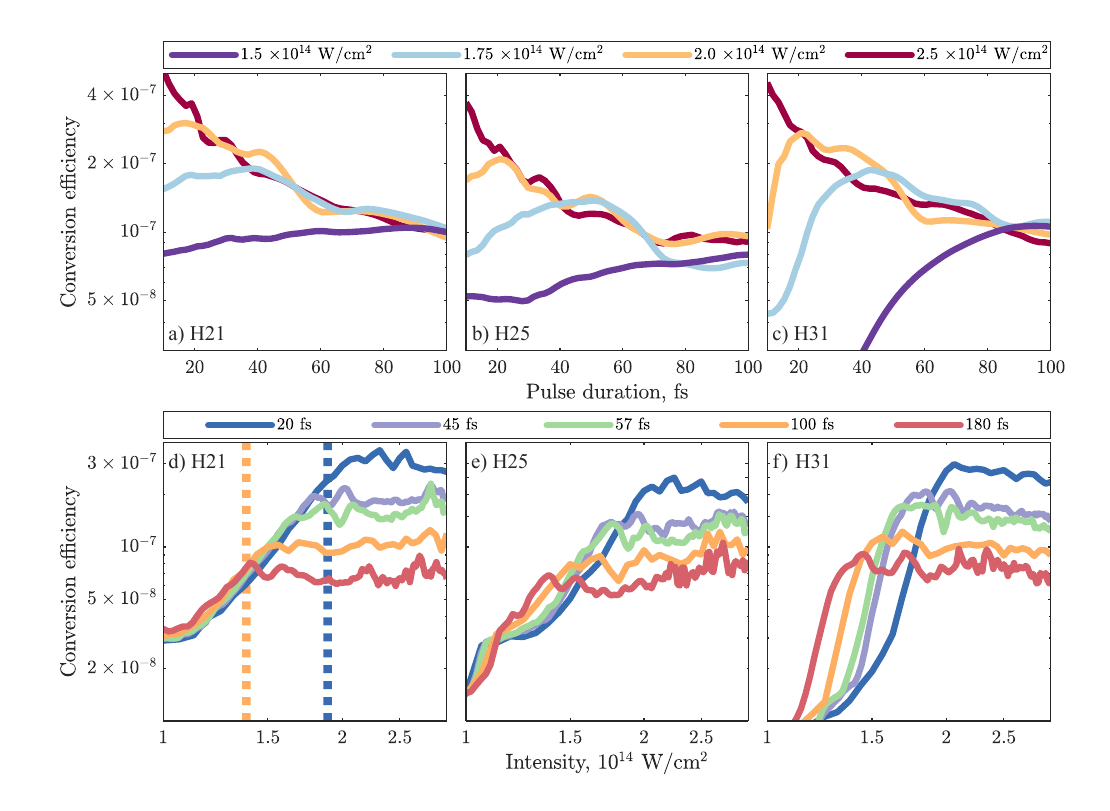}
    \caption{a)-c) Simulated CE for harmonics 21, 25 and 31 as a function of the pulse duration for different laser intensities. d)-f) Simulated CE for harmonics 21, 25 and 31 as a function of the laser intensity for different pulse durations. The dashed lines in d) represent the intensity at which critical ionization is reached for $\SI{20}{\femto\second}$ (blue) and $\SI{100}{\femto\second}$ (yellow).}
    \label{fig:SIM_21_25_31}
\end{figure*}

\subsection{Experimental results}

Figure~\ref{fig:Result_1} a) shows the total harmonic yield normalized by the energy of the generating field, ($\propto C_q$), as a function of pulse duration. The results obtained for different pulse energies are represented in color. Shorter driving pulses lead to a much higher conversion efficiency. There is almost one order of magnitude increase in normalized yield when using a $\SI{45}{\femto\second}$ pulse, compared to a $\SI{180}{\femto\second}$ pulse, in spite of a much lower pulse energy ($\SI{130}{\micro\joule}$ compared to $\SI{430}{\micro\joule}$). In the results of Fig.~\ref{fig:Result_1} a), we could not reach the optimum CE for all pulse durations, due to limited available pulse energy out of the MPC. Therefore, these results are only indicative of the variation of the CE as a function of pulse length.

Figure~\ref{fig:Result_1} b) shows the HHG spectrum corresponding to the three highlighted data points with durations $\SI{45}{\femto\second}$, $\SI{57}{\femto\second}$ and $\SI{180}{\femto\second}$. Shorter pulses lead to higher harmonic yields, slightly broader spectral widths, and higher photon energies. Indeed, atoms exposed to a shorter laser pulse experience a higher intensity when phase-matching is achieved since the critical ionization degree is reached at a higher intensity. This results in higher harmonic yields and an extended plateau. The increase of the harmonic bandwidth using short pulses can be understood by Fourier transform arguments and the effect of the intensity-dependent intrinsic harmonic chirp \cite{varju2004frequency}.

\subsection{Simulation results}

Figure~\ref{fig:SIM_21_25_31} a)-c) shows the simulated CE of harmonics 21, 25 and 31 in argon as a function of the pulse duration for four different laser intensities. At the highest intensity, the CE is the largest for the shortest duration for all harmonic orders. The CE reaches a maximum at $\SI{20}{\femto\second}$, $\SI{50}{\femto\second}$ and $\SI{100}{\femto\second}$, for intensities of  $2.0$, $1.75$ and \qtylist{1.5d14}{ W/c\square m}, respectively. 

Figure~\ref{fig:SIM_21_25_31} d)-f) presents the CE for the same individual harmonics as a function of laser intensity for five different pulse durations. The shortest pulse duration yields the highest CE for all harmonic orders at high laser intensity, while at low intensity the longer pulses have, in general, a higher CE. In all cases, there is an intensity at which the CE saturates, and as the pulse duration increases this saturation occurs earlier. The saturation effect happens when the critical ionization is reached, which takes place at increasing intensities for decreasing pulse durations. The vertical dashed lines in Fig.~\ref{fig:SIM_21_25_31} d) indicate these intensities for two durations, $\SI{20}{\femto\second}$ and $\SI{100}{\femto\second}$. The maximum CE is obtained for an intensity about $\SI{20}{\percent}$ above that indicated by the dashed lines. This can be understood by the fact that the optimum yield is achieved when phase-matching can be realized in a larger volume.

\subsection{Comparison between experiments and simulations}

\begin{figure*}
    \centering
\includegraphics[width=1\linewidth]{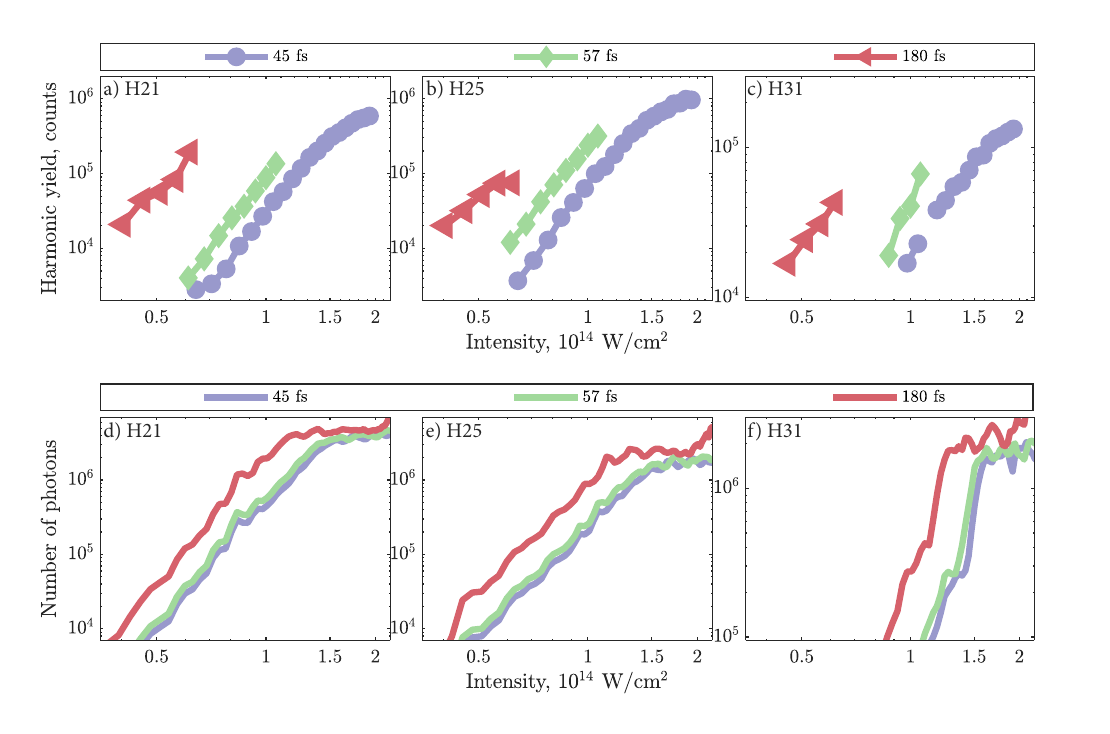}
    \caption{a)-c) Experimental yield of harmonics 21, 25 and 31 for pulse durations of $\SI{45}{\femto\second}$ (violet), $\SI{57}{\femto\second}$ (green) and $\SI{180}{\femto\second}$ (red). d)-f) Simulated number of photons for the same harmonics and pulse durations.}
    \label{fig:Theory_vs_experiment}
\end{figure*}

Figure~\ref{fig:Theory_vs_experiment} shows the experimental a)-c) and simulated d)-f) harmonic yields as a function of the laser intensity for harmonics 21, 25 and 31, at three pulse durations. Both experimental and simulated results exhibit the same behavior. To reach a given harmonic yield, higher intensities are required when using shorter pulses, but with much lower pulse energies. This leads to a shift in intensity between the harmonic yield curves corresponding to different pulse durations. This can be related to phase-matching being achieved at different critical intensities. The shift is slightly larger in the experimental results, possibly due to the difficulty in estimating intensities. 

\begin{figure}
\centering
    \includegraphics[width=1\linewidth]{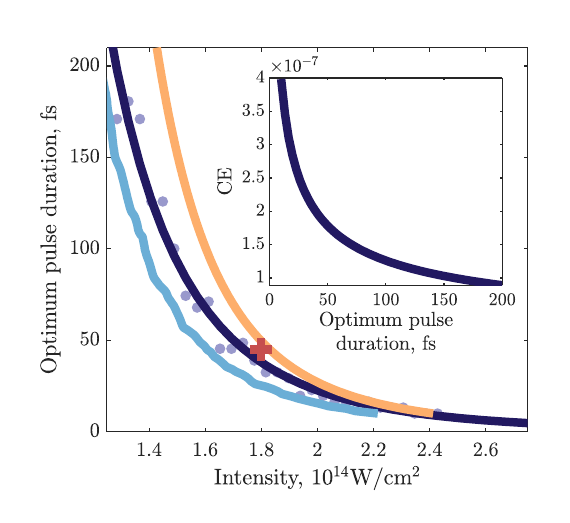}
    \caption{Pulse duration at which the CE of harmonic 25 is optimized as a function of laser intensity. The red cross at $\SI{1.8d14}{\watt\per\centi\square\meter}$ and $\SI{45}{\femto\second}$ is the optimum obtained from the experiment. Simulated data points are shown by violet dots, and a power fit of the simulated results is indicated by the dark blue curve. The yellow and light blue curves show the duration at which the critical ionization degree is reached at the peak of the pulse, obtained from the model presented in \cite{Minneker} 
    (yellow) and using ionization rates from TDSE (light blue). The inset shows the maximum CE as a function of the optimum pulse duration $\tau_\textrm{opt}$.}
    \label{fig:CE_sim}
\end{figure}

Finally, Fig.~\ref{fig:CE_sim} displays the duration that optimizes the CE for harmonic 25 as a function of intensity. The experimental result at $\SI{45}{\femto\second}$ is marked by the red cross, while simulated data points are marked by the violet dots. A power fit of the simulated results is indicated by the dark blue curve. The maximum CE is shown as a function of pulse duration in the inset. The corresponding dependence is approximately $\tau_\textrm{opt}^{-0.5}$, extracted from a power function fit. The yellow and blue curves represent the pulse duration at which the ionization degree, corresponding to perfect phase-matching, is reached at the peak of the pulse. The light blue curve uses ionization rates calculated by solving the TDSE. The yellow curve is obtained by utilizing the model proposed in Minneker et al. \cite{Minneker}, based on the Ammosov-Delone-Krainov approximation \cite{ammosov1986tunnel} to tunnel ionization. The difference at low intensity can be attributed to the breakdown of the quasi-static approximation in this regime \cite{PhysRevA.93.023413}. The slightly shorter estimated pulse duration of the light blue curve can be corrected if we take into account volume effects, similar to the difference between the dashed line and onset of saturation in Fig.~\ref{fig:SIM_21_25_31} d).

Extrapolating the dark blue curve in Fig.~\ref{fig:CE_sim} to a pulse duration of $\SI{40}{\pico\second}$, we find that the required laser intensity to achieve phase-matched HHG is of the order of $5\times10^{13}$ W/cm${^2}$, in good agreement with the experimental results presented in \cite{MFerray_1988}.
In the case when phase-matching does not depend on the ionization degree ($|\Delta k_{\text{fe}}|$ small), the CE is always optimized for the shortest pulse duration, independently of the laser intensity, leading to a horizontal line in Fig.~\ref{fig:CE_sim}.

\section{Conclusions}

We have developed a fs laser system with tunable Fourier-limited pulse duration ranging from $\SI{40}{\femto\second}$ to $\SI{180}{\femto\second}$, using an industrial Yb laser in conjunction with an MPC-based post-compression technique. This allows us to experimentally investigate the combined effects of the laser pulse duration and intensity on the generation of high-order harmonics in argon, thereby enhancing our overall understanding of the HHG process. We also perform extensive numerical simulations of HHG under the experimental conditions. 

Our main finding reveals that, for a given laser intensity, there exists an optimum pulse duration $\tau_\textrm{opt}$ that maximizes the CE. This suggests that long pulses are also suitable for phase-matched HHG, albeit with lower maximum CE and higher pulse energy requirements. The maximum CE that can be obtained varies as $\tau_\textrm{opt}^{-0.5}$. The results underscore that the HHG conversion efficiency is influenced by the interplay between the light-matter interaction and phase-matching of the generated fields in the non-linear medium. While the single-atom response increases with the laser intensity, the phase-matching condition imposes an intensity limit based on the required ionization degree, which is dependent on the pulse duration. 

\section*{Fundings}

The authors acknowledge the financial support from the Swedish Research Council (Grants No. 2021-04691 and 2022-03519), the European Research Council (Advanced grant QPAP), the Knut and Alice Wallenberg Foundation and the Crafoord Foundation. A.L. acknowledges support from the Knut and Alice Wallenberg Foundation through the Wallenberg Centre for Quantum Technology (WACQT). S.W. and G.B. acknowledge the support of the Helmholtz-Lund International Graduate School (HELIOS, project number HIRS-0018). M.A.V. acknowledges funding from the Italian Ministry of Research (MUR) via the ``I-PHOQS'' Grant (CUP B53C22001750006) and the COST Action CA18222 ATTOCHEM, supported by COST (European Cooperation in Science and Technology).

\begin{acknowledgments}

We acknowledge G. Arisholm for the theoretical support with simulations and the use of SISYFOS. 

\end{acknowledgments}

\section*{Disclosures}

The authors declare no conflicts of interest.

\section*{Data Availability Statement}

The data that support the findings of this study are available from the corresponding author upon reasonable request.

\section*{References}

\bibliography{aipsamp}

\end{document}